\documentclass[prb,aps,twocolumn,groupedaddress,floats,showpacs,final]{revtex4}
\usepackage{graphicx}
\usepackage{dcolumn}
\usepackage{bm}
\usepackage{color}
\usepackage{ulem}
\definecolor{blue}{rgb}{0.3,0.3,0.9}

\def\beq{\begin{eqnarray}}
\def\eeq{\end{eqnarray}}

\begin{document}

\author{Max Yarmolinsky and Anatoly Kuklov}
\address{$^1$ Department of Engineering \& Physics, CSI, and the Graduate Center of CUNY, New York.}
%\address{$^2$ Department of Physics, Odessa State University, Ukraine}

%\affiliation{Department of Engineering Science and Physics, CSI, CUNY, and The Graduate Center of CUNY}

\title{Revisiting universality of the liquid-gas critical point in 2D}

\date{\today}
\begin{abstract}
Critical point of liquid-gas (LG) transition does not conform with the paradigm of spontaneous symmetry breaking because there is no broken symmetry in both phases. 
 We revisit the conjecture that this critical point belongs to the Ising class by performing large scale Monte Carlo simulations in 2D free space in combination with the numerical flowgram method.
 Our main result is that the critical indices do agree  with the Onsager values within the error of 1-2\%. This significantly improves the accuracy reported in the literature. 
The related problem about the role of higher order odd terms in the (real) $\varphi^4$ field model as a mapping of the LG transition is addressed too. 
The scaling dimension of the $\varphi^5$ term at criticality is shown to be the same as that of the linear one $\varphi$. 
 We suggest that the role of all higher order odd terms at criticality is simply in generating the linear field operator with the critical dimension consistent with the Ising universality class.  
\end{abstract}

%64.70.Tg Quantum phase transitions
%05.30.-d Quantum statistical mechanics 
%05.50.+q Lattice theory and statistics (Ising, Potts, etc.)
%75.10.-b General theory and models of magnetic ordering

\maketitle

\section{Introduction}
The liquid-gas phase transition is characterized by latent heat which vanishes at one point of the phase diagram -- the critical point. Within the mean field approach, the LG coexistence curve is well described by the celebrated van der Waals equation  where the role of the order parameter is played by the difference in densities of liquid and gas (see in Ref.~\cite{Landau}).  Formally speaking, however, neither liquid nor gas can be characterized by a symmetry breaking order parameter simply  because there is no order in both phases.   

 Absence of any underlying symmetry breaking raised the question about the universality of the transition at the critical point. The standard conjecture is that this transition belongs to  the $Z_2$ universality class, that is, of the Ising transition (see in Refs. \cite{Landau,Kadanoff,Patashin}). 
This question have a straightforward answer for the lattice gas where a direct mapping to the Ising model exists \cite{Lee}. It is formally possible to consider a free space fluid on a lattice with spacing being much smaller than any typical distance determining  interaction. In this case the lattice and free space models should be equivalent. Thus,  in general, no underlying $Z_2$ symmetry can be found in such a lattice. Accordingly, lattice models explicitly violating $Z_2$ symmetry have been considered \cite{Mermin_PRL}. 
 It was further suggested that the asymmetry does not change the $Z_2$ universality of the LG criticality, and its role  is reduced to mixing of the primary scaling operators which results in the non-analytical corrections to the position of the critical point \cite{Mermin,Pokrovskii,Patashin,Bruce,Wilding}.  The extended mixing scenario has been suggested in Ref.\cite{Fisher2000,Fisher_2001,Fisher_2003,Fisher_2003_2} in relation to the  Yang-Yang anomaly.

 The conjecture  that LG criticality is $Z_2$ is closely related to the question about the role of higher order odd terms in the field theory. As shown in Ref. \cite{Hubbard}, the LG transition characterized by quite generic two-body interactions in free space can be mapped on a field theory of a continuous scalar real field    $\varphi$ with some effective Hamiltonian which, in addition to even terms $(\vec{\nabla} \varphi)^2,\, \varphi^2, \, \varphi^4,...$, contains 
odd ones $\varphi^1, \, \varphi^3, \, \varphi^5,\, \varphi^7, ...$. 
Thus, there is a possibility that higher order odd  terms $\varphi^5,\, \varphi^7, ...$  change the universality (the term $\varphi^3$ can be eliminated by a uniform shift $\varphi \to \varphi + \varphi_0$ with $\varphi_0$ being some constant) \cite{SMA}.  
The analysis \cite{Hertz} based on the renormalization group (RG) approach  found that there is a novel fixed point in dimensions $d=10/3$ induced by the term $\varphi^5$, provided, $\varphi^1 $ and $\varphi^3$ are tuned to zero.
This result, however, was challenged in Ref. \cite{Nicoll} based on the $\varepsilon$-expansion around $d=4$ showing that all odd operators of higher order are strongly irrelevant at the symmetric fixed point, so that this point is stable with respect to the odd perturbations. 

It is important to note that the argument \cite{Nicoll} cannot be used in 2D. Thus, the question about the role of the higher odd terms in 2D remains open. More recently, the analytical solution for the critical exponents of 3D LG transition has been found under quite general assumptions \cite{Bondarev}. These exponents turn out to be different from the values obtained numerically. The same method can also be used in 2D and it gives the exponents which are different from the Onsager values \cite{Bondarev_priv}.

 Some early attempts to measure critical exponents experimentally have claimed significant deviations from the 3D Ising universality \cite{Wulkowitz,Garland}, while others \cite{Hayes,Moses} find an acceptable agreement with the Ising universality, provided the fitting procedure included subcritical corrections (with several adjustable parameters). The main problem turns out to be due to gravity which does not allow to approach the critical point close enough so that the corrections to the leading scaling can be ignored.
The  experiments in microgravity (see in Ref. \cite{RMP_2007}) didn't improve the situation much. 

Measurement of the LG criticality in 2D has been conducted in Ref.\cite{Moses_2D}. The value of the $\beta$-exponent was reported to be consistent with the Onsager result $\beta=1/8$ within 15-20\% accuracy. This result was achieved within 3-parametric fitting procedure requiring knowledge of accurate values of the critical temperature and density. At this point we note that the value of $\beta =1/8$ is also characterizing other universalities, e.g.,  XY and three-state Potts model. Thus, by itself it is not a "smoking gun" for the Ising criticality.

The LG critical point has been addressed by direct Monte Carlo simulations by many groups. In Ref.\cite{Bruce} the analysis of 2D Lennard-Jones fluid has been carried out within the hypothesis of the mixing \cite{Mermin,Pokrovskii,Patashin}, and it has been concluded that the universality of the transition is consistent with the Ising class. However, the maximum size simulated in this work allowed to include only about 400 particles on average, with two relatively small sizes of the simulation box used. Under this condition the applicability of the finite-size scaling (FSS) analysis becomes questionable. The same approach has been used in 3D \cite{Wilding} with the conclusion that the 3D LG critical point belongs to the $Z_2$ class. The role of corrections to scaling turns out to be much more important in 3D. This, in particular, lead to inconsistent values of the $\nu$ exponent deduced from different quantities.       

Monte Carlo simulations have been also conducted  for the model interaction potential -- the square well in 3D in Ref.\cite{SW} (see also references there).  The analysis was carried out for a set of box sizes from 6 to 18 hard core radii, and the conclusion was reached that the universality of the critical point is consistent with the Ising class. Later, however, a different result has been obtained for Lennard-Jones potential \cite{Pap} -- the critical exponent $\nu$ was not consistent with the Ising class. 
The LG criticality has been also addressed in a series of papers \cite{Fisher_2001,Fisher_2003,Fisher_2003_2}, where both the critical exponent $\nu$ and the critical histogram were found to be consistent with those of the 3D Ising. [At this point, however, we should notice that the accuracy in the $\nu$-exponent value does not allow to exclude the non-Ising universality \cite{Bondarev}]. 
The approach based on molecular dynamics has been utilized in Ref.\cite{Watanabe} and significantly larger sizes have been simulated with the conclusion that the LG criticality in 3D is of Ising type.

It is important to note that the methods used to evaluate the critical exponents in Refs. \cite{Bruce,Wilding,SW,Pap,Watanabe} are strongly dependent on the choice of the values of the critical temperature $T_c$ and pressure $P_c$ (or density). This introduces significant uncertainties in the exponents.  In 3D the corrections to scaling must also be included. Thus, the fits become multi-parametric which introduces even larger errors. Furthermore, as pointed out in Ref.\cite{Fisher2000}, the Yang-Yang singularity
implies non-analytical corrections to the position of the  liquid-vapor  coexistence line which makes questionable the extrapolation procedures for the purpose of recovering the $\beta$ exponent.  
 %     Indeed, if temperature $T$ approaches its critical value $T_c$, the error in the exponent $\nu$ determining the correlation length $\xi \sim |T-T_c|^{-\nu}$ scales as $\delta \nu /\nu \sim \delta T_c/|T-T_c|$, where $\delta T_c$ is an error in determining $T_c$. Thus, inaccurate, say, third digit in $T_c$ can produce $\sim 100$\% error if $|T - T_c|/T_c \sim  0.01$. Since more accuracy in the power law fit requires closer approach to $T_c$, the statistical data must be collected with accuracy determined by the ratio $|T - T_c|/T_c$. This makes the fit procedure very challenging.   

Overall, it is fare to say that the majority in the scientific community does accept the conjecture that the LG criticality belongs to the Ising class despite that the experimental and numerical evidence may leave some room for doubts  due to substantial uncertainties in measured indices. Thus, our main motivation is to significantly improve the accuracy in determining the indices.
Here we suggest a different approach -- based on the so called {\it numerical flowgram} (NF) first introduced in Ref.\cite{Annals} and further developed in Ref.\cite{NJP}. This method is based on the finite size scaling (FSS) approach \cite{FSS}. It allows finding  position of the critical point as a byproduct of tuning a system into criticality with the help of the Binder cumulants \cite{Binder}.  Thus,  error in the critical exponents is given essentially by the error of the Binder cumulant only --  because no extrapolation or multi-parametric fit procedure are used. 

We apply the NF method to the LG critical point in 2D by measuring directly the critical index $\mu$ (and, independently, $\gamma/\nu$ as a crosscheck). The outcome of our large scale simulations allows to conclude with high certainty that the 2D LG criticality does belong to the Ising class.
 It is important to note that our analysis is not affected by the mixing effect. 
  Using the same method we have determined the scaling dimension $\Delta_5$ of the $\phi^5$ term in the $\varphi^4 + \varphi^6$ model in the context of the correspondence between the LG and the field ensembles. Our finding is that $\Delta_5$  coincides with that of the linear term in the $Z_2$ class.  

Our paper is organized as follows. First we address the role of the odd term $\varphi^5$ in the mapping \cite{Hubbard} of the LG criticality to the field theory in 2D. Then, we present the results of the direct simulations of the LG critical point in 2D. These parts are independent from each other with the exception that the same NF method is implemented for both. 
Finally, we discuss the results and open problems and outline a path toward detecting the non-analyticity induced by the Yang-Yang anomaly \cite{Fisher2000} within the NF approach.

\section{Critcality with the $\varphi^5$ term}
As discussed above, there is a formal mapping between a gas of particles undergoing the LG transition and the field theory \cite{Hubbard}. This mapping, however, unavoidably contains odd terms in the field.  
The proposal \cite{Hertz} of the asymmetric fixed point is based on the assumption that the operator $Q_5 = \int d^dx \varphi^5$ in the field model is relevant at the symmetric fixed point in $d<10/3$ -dimensional space. Then, the symmetric point may become unstable and the system finds another (asymmetric) fixed point characterized by critical indices different from those of the Ising model \cite{Hertz}. 
The alternative view based on the $\epsilon$-expansion around $d=4$  renders $Q_5$ and all higher terms as (dangerously) irrelevant \cite{Nicoll}. This argument, however, cannot be used in 2D. Thus, the issue of the odd terms remains quite controversial in 2D, and our goal here is to resolve it by simulations.

Here we will specifically focus on the critical dimension $\Delta_5$ of the $Q_5$ term in the potential part of the action $V(\varphi)$ characterized by the symmetry $\varphi \to - \varphi$. At this point it is important to mention that the result of adding $g_5 \varphi^5 $ to $V(\varphi)$ can be quite drastic at the microscopic level already -- this term can simply eliminate the transition before scaling behavior develops. We are not considering this option, and focus on the situation where $Q_5$ term is small at the microlevel. Then, if it is relevant in the sense of renormalization, it will take  the system away from the Ising fixed point to a new (non-Ising) one.

  At this point it is important to realize that the paradigm of {\it universality} implies that the microscopic form of the action $V(\varphi)$ does not affect the scaling behavior occurring around $\varphi=0$. The only requirement is that this action should have
 not more than two equilibrium solutions in the vicinity of $\varphi=0$ away from the critical point. Traditionally, the action is taken as a truncated polynomial $V(\varphi)=\sum_{n=1}^{n*} g_{2n} \varphi^{2n}$ with $n^*$ being as small as possible to insure overall stability. In the presence of the $\varphi^5$ term, $n^*=3$ is sufficient.   Thus, a natural choice of the model corresponds to the uniform part of the action $H_u=\int d^dx [V(\varphi) - g_5 \varphi^5]$ with 
\beq
V(\varphi) =g_2 \varphi^2 + g_4 \varphi^4 + g_6 \varphi^6,
\label{phi6}
\eeq 
where $g_2,\, g_4 >0,\, g_6>0, \,g_5$ are parameters. Without loss of generality we will be using $g_4=g_6=1,\, g_5>0$.
The range of values of $g_5$ is chosen in such a way as to avoid creating  extrema additional to $\varphi=0$ -- at least at the mean field level. This corresponds to the condition 
\beq
|g_5| < g^*= \frac{16}{5\sqrt{3}}\sqrt{g_4g_6}\approx 1.848  
\label{g5}
\eeq
 implying that the $Q_5$ term does not disturb the system strongly at the microscopic scale. Fluctuations may change this situation. 
Thus, in simulations we will consider the range $0< g_5 < g^*$. 
According to the standard practice \cite{Landau}, the action (\ref{phi6}) must be supplemented by the gradient term $ \sim \int d^dx  (\vec{\nabla}\varphi)^2 >0$. 

Simulations have been conducted in 2D for the discretized version of the model -- placed on a square lattice.
Then, the partition function becomes
\beq
Z=\int D\varphi \exp(-H),
\label{Z66}
\eeq
with
\beq
H= -t \sum_{\langle ij\rangle} \varphi_i \varphi_j  + \sum_i [V(\varphi_i) - g_5\varphi^5_i],
\label{H66}
\eeq
where the field $\varphi_i$ is defined at a site $i$ of the square lattice with $L$ sites along each direction, and the summation $ \sum_{\langle ij\rangle}$ runs over nearest neighbor sites separated by $\Delta L=1$ distance and coupled by the parameter $t>0$. This parameter together with $g_5$ will be used to tune the system into the critical point. Thus, in addition to $g_4=g_6=1$ we set $g_2=1$. The measure in (\ref{Z66}) is defined as $\int D\varphi = \prod_{i=1}^{L^2} \int_{-\infty}^{\infty} d\varphi_i$.

We will be using the dual formulation of the model (\ref{Z66},\ref{H66}) in terms of the non-oriented loops and will utilize the Worm Algorithm \cite{WA}. More specifically,  the factor $\exp(t \varphi_i \varphi_j)$ at each bond as well as $\exp(g_5 \varphi_i)$ at each site  are expanded in Taylor series and, then, each term is integrated out with respect to the field $\varphi_i$. The resulting partition function (\ref{Z66}) is represented in terms of the powers and coefficients of the expansion as
\beq
Z=\sum_{\{N_{ij}\}, \{n_i\}} \prod_{\langle ij\rangle}  \left(\frac{t^{N_{ij}}}{N_{ij}!}\right) \prod_i \left(S(C_i) \frac{g_5^{n_i}}{n_i!}\right),
\label{ZZ6}
\eeq
 where $N_{ij}=0,1,2,..., \infty$ are integers defined at bonds between neighboring sites $i$ and $j$; $n_i=0,1,2,..., \infty$ are defined at sites,  and
\beq
S(C_i) = \int_{-\infty}^\infty d\varphi \varphi^{C_i} \exp(- a\varphi^2 - g_4\varphi^4 -g_6\varphi^6),
\label{S6}
\eeq
with
\beq
 C_i =\sum_{j=<i>} N_{ij} + 5n_i ,
\label{C6}
\eeq
where $\sum_{j=<i>}$ denotes summation over bonds connected to the site $i$. Thus, the configurational space is fully defined by
the bond and the site integers $N_{ij}, n_i$, respectively .

The inspection of Eq.(\ref{ZZ6}) indicates that the partition function can be represented as a series in {\it even} powers of $g_5$:
\beq
Z=\sum_{N_5=0,2,4,...} B_{N_5}\cdot g_5^{N_5},\,\,\,\, N_5 = \sum_i n_i,
\label{N5}
\eeq
where
\beq
B_{N_5}=\sum_{\{N_{ij}\}}\sum_{\{\sum_i n_i=N_5\}} \prod_{\langle ij\rangle}  \left(\frac{t^{N_{ij}}}{N_{ij}!}\right) \prod_i \left( \frac{S(C_i)}{n_i!}\right)
\label{CN5}
\eeq
are positive coefficients independent of $g_5$. 
This is consistent with the symmetry of the model with respect to simultaneous change $\varphi  \to - \varphi,\,\, g_5 \to - g_5$.     
Thus, the dual representation (\ref{ZZ6}-\ref{C6})  is free from the sign problem.

While being formally exact in the asymptotic sense, the mapping of the LG transition on the field theory \cite{Hubbard} is not practical for obtaining specific results if viewed beyond the paradigm of universality -- simply because the resulting action is presented as an infinite series. Thus, the analysis of a field model in conjunction with the LG criticality makes only sense along the line of the universality concept when the action is truncated.  To emphasize this aspect
 we introduce a variety of models which, despite having very different appearance, demonstrate the same critical behavior.   

It is also useful to use a simplified (for numerical purposes) version of the model -- by limiting the onsite values of $n_i$ in Eqs. (\ref{ZZ6},\ref{N5}) to
$n_i=0,1$ only.  In other words, in the expansion of $\exp(g_5 \varphi^5_i)$ in Eq.(\ref{Z66},\ref{H66}) only  two first terms are kept. According to the paradigm of universality such a truncation should not affect the scaling properties of the model -- that is, in the limit when the correlation length exceeds considerably the lattice constant. This truncation corresponds to the partition function    
\beq
Z=\int D\varphi \exp(-H_1)\prod_i (1+g_5 \varphi^5_i),
\label{Z}
\eeq
where
\beq
H_1= -t \sum_{\langle ij\rangle} \varphi_i \varphi_j  + \sum_i [ a \varphi_i^2 + g_4 \varphi_i^4 +g_6\varphi^6_i] .
\label{H}
\eeq
 Following the standard approach \cite{Landau} that only the first most relevant terms of the Landau expansion matter, the integrand in Eq.(\ref{Z}) can be rewritten as $\exp(-H_1 +\ln(1+g_5\varphi^5)) \to \exp(-H_1 +g_5\varphi^5 -  g_5^2\varphi^{10}/2 )$, with the higher order terms dropped. As it is obvious, the truncated model does not need to have the $\sim g_6$ term because there is no instability anymore -- due to the term $\sim \varphi^{10}$.     
Thus, $g_6$ can be set to zero in Eq.(\ref{H}). 

A comment is in order about the appearance of the model (\ref{Z}) which may invoke the sign problem  because the integrand in Eq.(\ref{Z}) is not positively defined. As clearly seen from the representation (\ref{N5}) valid for both models, each term in the series is positive, and, thus, there is no sign problem in the truncated model as well. In principle, one can generate arbitrary number of the truncated models which are free from the sign problem --
by limiting the onsite factors $n_i$ up to some maximum value greater than 1. This limitation, obviously, should have no impact on the scaling behavior.

The dual representation (\ref{ZZ6}-\ref{C6},\ref{N5},\ref{CN5})) is especially convenient in calculating the mean thermodynamical values $\langle ...\rangle$ of  $\sum_i \varphi^5_i$. 
 Evaluation of $ d \ln Z/dg_5$ in the representations 
(\ref{N5}) and (\ref{Z66}) gives 
\beq
\langle \psi \rangle = g_5^{-1} \langle N_5\rangle, \,\, \psi= \sum_i \varphi^5_i .
\label{psi}
\eeq
Similarly, higher order means $\langle \psi^m \rangle$, $m=2,3, ...$  can be expressed in terms of the means  of the higher powers of $N_5$.

 For the truncated model, the derivative $ d \ln Z/dg_5$ applied to the representation (\ref{Z}) and compared with (\ref{N5}) gives the relation similar to Eq.(\ref{psi}):
\beq
\langle \psi_1 \rangle = g_5^{-1} \langle N_5\rangle, \, \psi_1= \sum_i \frac{\varphi^5_i}{1+g_5 \varphi^5_i} \to \psi,
\label{psi1}
\eeq
where the last relation is written with respect to the limiting scaling behavior. This aspect will be explicitly addressed below.

The paradigm of Universality predicts that both models should have the same critical behavior.
We will present results of the simulations for the truncated as well the full model. Jumping ahead, it will be shown that, while the position of the critical point, $t=t_c$, is different for two models, the critical behaviors are identical within the statistical error (of about 1-2\%).  

It is important to report that we have found no fixed point  at any finite value of $g_5$ within the interval $0<g_5 \leq 1$ (where the correlation length is diverging). Thus, we conclude that there is only one fixed point -- corresponding $g_5=g_c=0$. Then the question should be answered about the scaling dimension $\Delta_5$ of the $g_5$-term. 
This can be achieved by observing the divergence of the correlation length $\xi \sim g_5^{-\mu_5}$ with some exponent $\mu_5>0$ as $g_5 \to 0$ as long as $t=t_c$. Such a divergence has been observed and it is found that $\mu_5$ coincides with the Onsager value $\mu=8/15$ of the field exponent (within 1-2\% of the total error). This implies that $\Delta_5=2-1/\mu=1/8$ is the same as the critical dimension $\Delta_1$ of the field $\varphi$.

\subsection{Critical behavior at $g_5=0$ by the flowgram method}
The idea of the flowgram method \cite{Annals,NJP} is based on constructing the FSS flow (with respect to the system size $L\to \infty$) by adjusting a critical parameter  $t$ so that some Binder cumulant $U_B$ \cite{Binder} is tuned to a value within  its critical range. Conversely, keeping $U_B$ within its critical range (by adjusting $t$) as $L\to \infty$ guarantees that $t \to t_c$ with  increasing accuracy. Then, a quantity $Q$ characterized by scaling  behavior will exhibit self-similar dependence versus $U_B$  with respect to $L$. In other words, if $U_B$ is kept in the critical range for large enough $L$, the plot  $Q$ versus $U_B$ can be represented by some universal function multiplied by the factor $L^{-\Delta_Q}$ with the exponent $\Delta_Q$ determining scaling dimension of $Q$.  

More specifically, far from the criticality $U_B$ takes some fixed values, say, $U_B=B_0$ in the disordered phase and $U_B=B_1$ in the ordered phase. At the critical point, $t=t_c$ (and $g_5=0$), it takes a value $U_B=B_c$ independent of the system size $L$ as long as $L \to \infty$ and such that $B_0 < B_c<B_1$ ( where for the sake of argument we assume $B_1>B_0$).
It is important to note that for any finite $L$ the function $U_B(t)$ changes smoothly from $B_0$ to $B_1$ as $t$ passes from $t<t_c $ to $ t>t_c$. However, as $L$ is taken larger and larger, the domain $\delta t$ around $t=t_c$ over which this change happens becomes smaller and smaller. Thus, in the thermo-limit ($L \to \infty$) the cumulant exhibits a jump from $B_0$ to $B_1$
at exactly $t=t_c$ because $\delta t \sim L^{-1/\nu}$  in accordance with the FSS \cite{FSS}, with $\nu >0$ being the critical exponent characterizing the divergence of the correlation length $\xi \sim |t-t_c|^{-\nu}$. 

This strategy is guaranteed to access a critical point in progression of growing sizes $L$ -- as long as $U_B$ is tuned to any value within the critical range $B_0< U_B<B_1$. Accordingly, the system is always in the critical range of $U_B$ (and of any other scaling quantity). In particular, the family of curves $ dU_B/dt$ vs $U_B$ for various $L$ must be self-similar for large enough $L$ because
$dU_B/dt \approx (B_1-B_0)/\delta t \propto L^{1/\nu}$.   Thus, constructing such a family and then rescaling them into a single master curve by a scaling factor $\lambda(L)$ gives the exponent $\nu$ by plotting $\ln \lambda $ vs $\ln L$. Similarly, other exponents can be found by choosing the appropriate quantity $Q$ to plot versus $U_B$ and to perform the rescaling of the family of the curves (for various $L$) into a single master curve. Clearly, within this approach the value of $t_c$ plays no explicit role in the fitting procedure, with the only one fitting parameter being the scaling dimension.
\begin{figure}
\vspace*{-0.5cm}
 \includegraphics[width=1.1 \columnwidth]{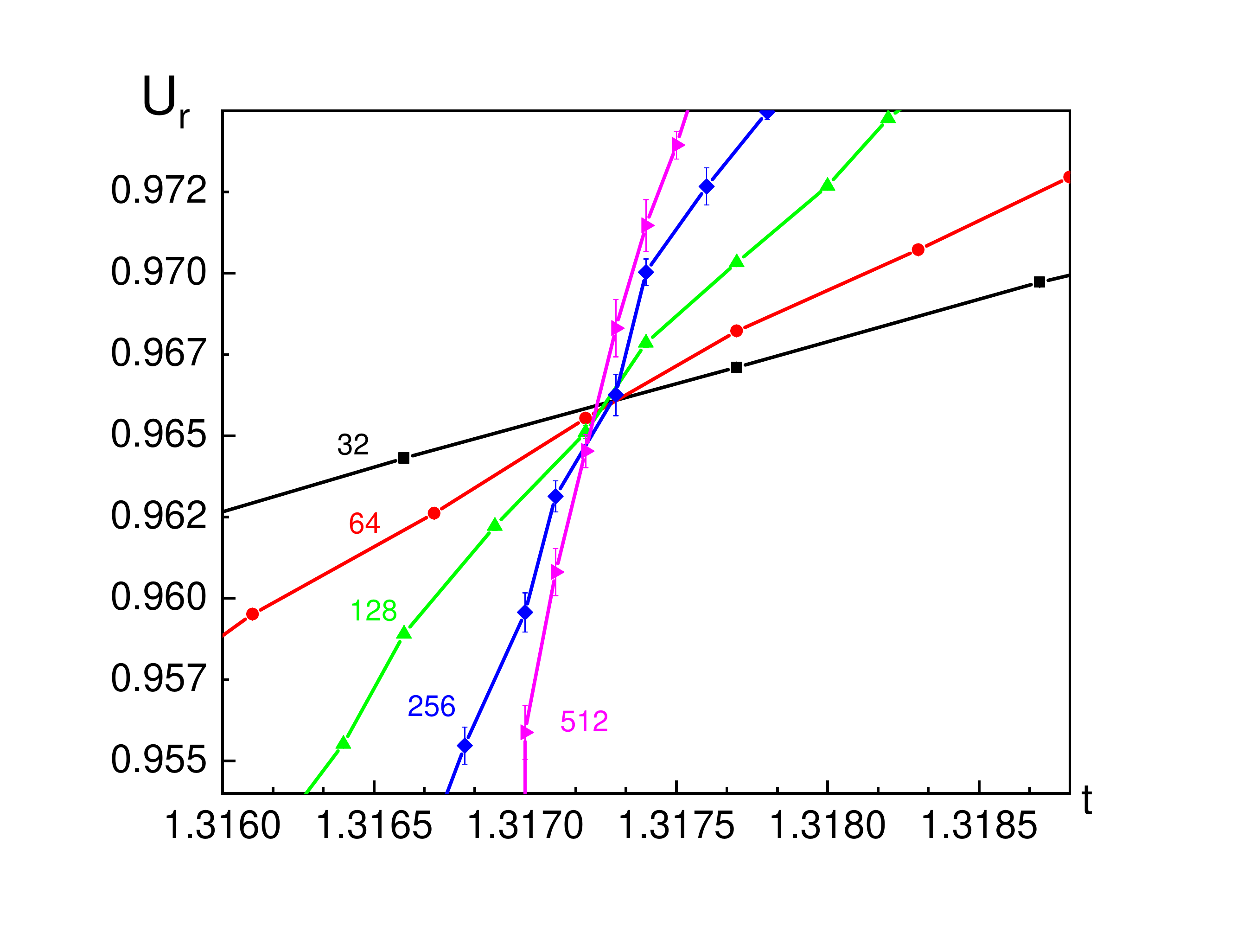}
\vskip-8mm
\caption{(Color online) $U_r$ vs $t$ for various $L$ (shown close to each curve); lines are guides to eye.The crossing point corresponds to $U_B=0.965$ and   it determines $t_c =1.3173 \pm 0.0003$ (for $g_4=1,\,g_6=0,\, g_2=1, \, g_5=0$ in Eqs.(\ref{ZZ6},\ref{H66})).}
\label{fig:cross}
%\vskip-5mm
\end{figure}
\begin{figure}
\vspace*{-0.5cm}
 \includegraphics[width=1.1 \columnwidth]{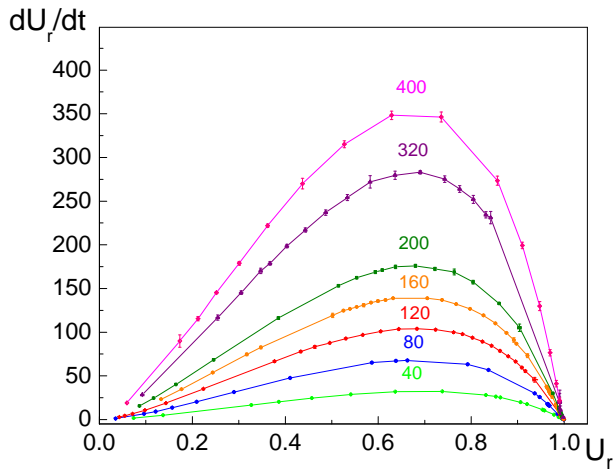}
\vskip-8mm
\caption{(Color online) Monte Carlo results for $dU_r/dt$ vs  $U_r$ as defined in Eq.(\ref{dUb}) for several system sizes $L$ shown close to and above each curve. Lines are guides to eye.}
\label{fig:dUB}
\vskip-5mm
\end{figure}
\begin{figure}
\vspace*{-0.5cm}
 \includegraphics[width=1.1 \columnwidth]{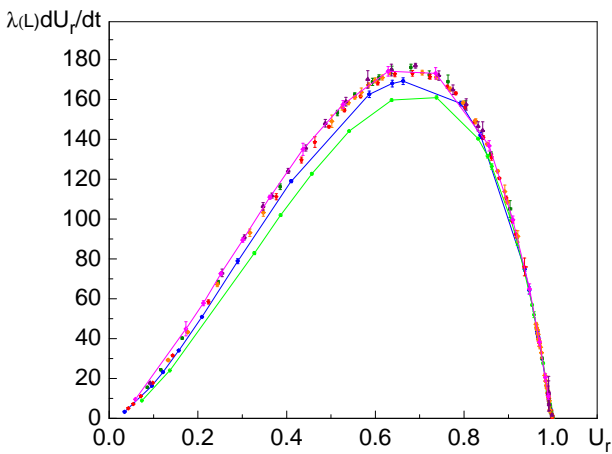}
\vskip-8mm
\caption{(Color online) Rescaled data shown in Fig.~\ref{fig:dUB} with $\lambda(L)=(200/L)^{1/\nu},\, \nu=1$. The overall statistical error of the data is $\sim 1-2\%$.}
\label{fig:master}
\vskip-5mm
\end{figure}

In order to determine the $\nu $ exponent we have chosen the following Binder cumulant
\beq
U_r(t,L)=\frac{\langle r^2\rangle_G }{r^2_L },\,\, r^2_L=\sum_{\vec{r}} \vec{r}^{\,2}/L^d \propto L^2,
\label{Ub}
\eeq
where $\langle r^2\rangle_G= \sum_{\vec{r}} G(\vec{r}) \vec{r}^2 /\sum_{\vec{r}} G(\vec{r})$, with  $G(\vec{r})$ denoting the correlator $\langle \varphi(\vec{r}) \varphi(0) \rangle$ taken at two points in 2D space separated by the vector $\vec{r}$; and $\langle ...\rangle $ defines the averaging with respect to the partition function (\ref{ZZ6}). 
To demonstrate that $U_r$ is a scale invariant quantity at the critical point, we have analyzed its behavior vs $t$ for various sizes. Fig.~\ref{fig:cross} shows the crossing point of $U_r$ at $t=t_c\approx 1.3173$ for the parameters $g_2=g_4=1,\, g_6=0,\, g_5=0$.  The value of $t_c$ depends on $g_6$. For the case $g_2=g_4=g_6=a=1,\, g_5=0$ it is $t_c \approx 1.6975$.
[The accuracy of $t_c$ is controlled by the maximum system size $L$ simulated].
  
By the definition, Eq.(\ref{Ub}),  $U_r\to 0$ (as $L\to \infty$) in the disordered phase (where the correlation length is $\sim {\cal O}(1)$) and $U_r=1$ in the ordered phase where the coherence length reaches the system size $L$. Thus, formally speaking, any value in the interval  $0< U_r<1$ belongs to the critical range of $U_r$.  In reality, for practical purposes of achieving better accuracy of the critical exponent we have found that it is reasonable to tune $U_r $ into the region where $dU_r/dt $ vs $U_r$ reaches its maximum (see  Fig.~\ref{fig:dUB}), that is, within the range $0.5 < U_r< 0.8$.

  At $g_5=0$ the integers $N_{ij}$ form closed non-oriented loops. Within the Worm Algorithm \cite{WA} the evaluation of the correlator corresponds to having one loop with two open ends. In this space, $U_r$ can be constructed as the histogram of the square of the distance $\vec{r}^{\,2} $ between two  open ends which represent two random walkers. Accordingly $dU_r/dt$ can be found as 
\beq
t \frac{dU_r}{dt}= \sum_{\langle ij \rangle} [\langle  N_{ij}  \vec{r}^{\,2}\rangle_G -  \langle  N_{ij} \rangle_G \langle \vec{r}^{\,2}\rangle_G]
\label{dUb}
\eeq
 following direct differentiation vs $t$ in the dual representation (\ref{ZZ6},\ref{S6},\ref{C6}) .

The result of this procedure -- the family of graphs $dU_r/dt$ vs $U_r$ for various $L$ is shown in Fig.~\ref{fig:dUB} for $g_4=a=1,\, g_6=0$. 
The master curve obtained by the vertical rescaling of the data with the exponent $\nu=1$ is shown in Fig.~\ref{fig:master}.
The lines connecting the data points  for $L=40,80$ are shown in order to emphasize that at these sizes the sub dominant term is still visibly significant so that these data points do not collapse into the master curve. The line for $L=400$ is also shown to indicate that all higher sizes $L=120,160,200,320,400$ belong to to the master curve within the error 1-2\%.   

At finite $g_5$ the structure of the configurational space changes -- there are loops which are not closed. The general condition (\ref{C6}) indicates that whenever $n_i=1,3,5$ at a site $i$, there is an odd total number of the  integers $N_{ij}$ at the bonds connecting this site with its neighbors $j$.

\subsection{Critical behavior at finite $g_5$}
Ising critical behavior is characterized by two primary fields $\sim \varphi^2$ and $\sim \varphi$ with the corresponding "charges" $\tau \sim t-t_c$ and $h$. In the space $(\tau,h)$ the divergence of the correlation length $\xi$ along the line $h=0$ is characterized by $\xi \sim \tau^{-\nu} \to \infty$ and by $\xi \sim h^{-\mu} \to \infty $ along $\tau=0$, with the Onsager exponents $\nu=1,\, \mu=8/15$. According to the FSS, once $\xi$ reaches the system size $L$, the role of $\xi$ is taken over by $L$. In the previous section we have explored the first property and have shown that the $\nu$ exponent is consistent with the Onsager solution. In order to observe the divergence along the second line one should select $t= t_c  $ as determined from the previous procedure for largest sizes and to apply the NF method --  now at finite $h$. In this case plotting $dU_B/dh$ vs $U_B$ for various $L$ and constructing the master curve by rescaling $dU_B/dh$ into a single master curve by some factor $\lambda(L)$ for each $L$ will give the $\mu$ exponent. 

The above logic can be followed in order to determine scaling dimensions of any higher odd terms. Here we will be concerned with the term $\sim \varphi^5$ as the most possibly relevant one -- as suggested in Ref.\cite{Hertz}. We have determined the corresponding critical exponent $\mu_5$ from the rescaling procedure of the graphs $ dU_B/dg_5$ versus $U_B$ for various $L$.

At thus juncture we have to change the type of the Binder cumulant $U_B$.
 At finite $g_5$ (or in the presence of any other odd term) using the cumulant $U_B=U_r$, Eq.(\ref{Ub}), is not convenient  because the number of open loops is now a dynamical variable. Thus, we choose
 $U_B=U_2= \langle \sum_i\varphi_i^5 \rangle^2/\langle (\sum_i\varphi_i^5)^2 \rangle $ built on the $\varphi^5$ term.
In the dual representation (\ref{ZZ6}) it is  %which is given by $dZ/dh$ and $d^2Z/dg^2_5$. 
\beq
U_2= \frac{(d\ Z/dg_5)^2}{Zd^2 \ln Z/dg^2_5}= \frac{\langle N_5 \rangle^2}{\langle N_5(N_5-1) \rangle}.
\label{U2}
\eeq     
For the full model (\ref{Z66},\ref{H66}) $U_2 =\langle \psi \rangle^2/\langle \psi^2 \rangle$, where $\psi $ is defined in Eq.(\ref{psi}).  
Clearly, $U_2=0$ at $g_5=0$ simply because $\langle \psi \rangle =0$ and $\langle \psi^2 \rangle$ is finite;  and $U_2=1$ far away from the critical point --  where  $g_5 \neq 0$ and fluctuations are suppressed. 
\begin{figure}
\vspace*{-0.5cm}
 \includegraphics[width=1.0 \columnwidth]{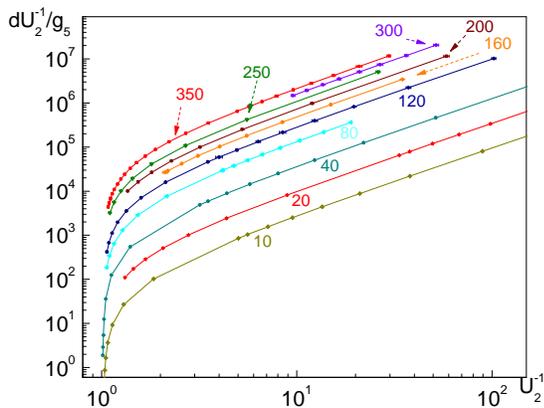}
%\vskip-8mm
\caption{(Color online) Monte Carlo results for $dU^{-1}_2/dg_5$ vs  $U^{-1}_2$ in the truncated model (\ref{Z})  for sizes $L$ shown close to each curve.}
\label{fig:dU2}
\vskip-5mm
\end{figure}
\begin{figure}
%\vspace*{-0.5cm}
 \includegraphics[width=1.0 \columnwidth]{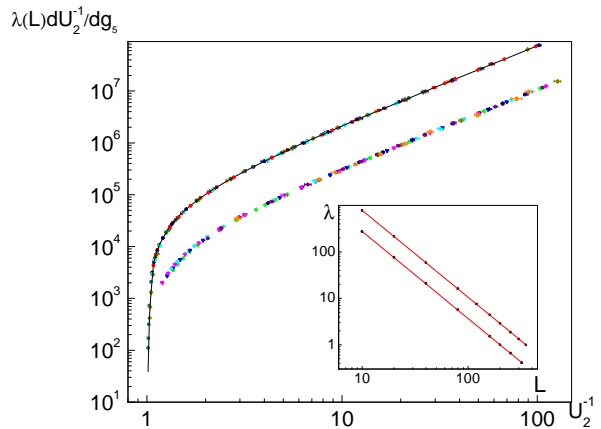}
\vskip-8mm
\caption{(Color online) The master curve obtained by "vertical" rescaling of the plots $dU^{-1}/dg_5$ vs $U_2^{-1}$. The upper curve is from Fig.~\ref{fig:dU2} obtained by the rescaling by the factor $\lambda(L)$ to match the data for $L=350$, that is, $\lambda(350)=1$. The lower curve is obtained by the same procedure for the data obtained from the full model, $g_6=1$, for $L=10,20,40,80,160,200,250,320$, with the choice $\lambda(200)=1$.   Inset: The log-log plots of $\lambda$ versus $L$ for the full (the lower data and the line) and truncated (the upper data and the line) models. Solid lines are the linear fits  with the slopes  $1/\mu_5$ giving $\mu_5=0.534 \pm 0.008$. The error includes statistical and systematic contributions. This value is consistent with the Onsager exponent $\mu=8/15\approx 0.533$.  }
\label{fig:masterg5}
\vskip-5mm
\end{figure}
\begin{figure}
%\vspace*{-0.5cm}
 \includegraphics[width=1.0 \columnwidth]{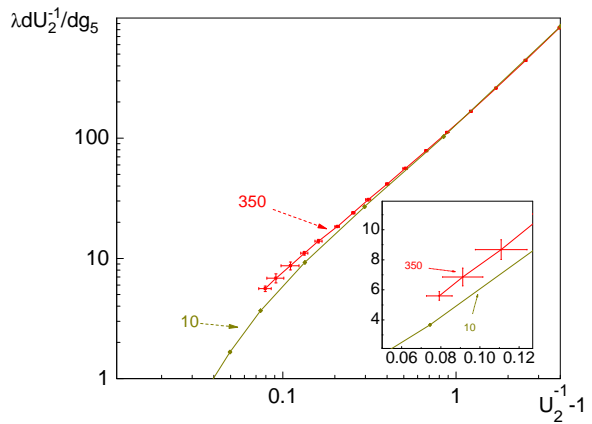}
\vskip-8mm
\caption{(Color online) Deviation from the scaling. Two curves, $L=10,350$, from Fig.~\ref{fig:masterg5} are shown in the domain where deviations from scaling are significantly higher than the statistical error of 1\% (about 15\%). Inset: More detailed view on the linear scale.}
\label{fig:deviation}
\vskip-5mm
\end{figure}
For the truncated model the role of $\psi$ is played by $\psi_1$, Eq.(\ref{psi1}). In the limit $g_5 <<1$ the denominator in $\psi_1$ plays no practical role.
More specifically for the truncated model  $\langle N_5(N_5-1) \rangle =g_5^2 [\langle \psi ^2\rangle - \sum_i \langle \varphi_i^{10}/(1+g_5 \varphi^5_i)^2\rangle] \to g_5^2 \langle \psi ^2\rangle$ because the term $\sim \langle \psi ^2\rangle$ has the extra factor $L^2$ with respect to $\sim \sum_i \langle \varphi_i^{10}\rangle $.
 We will be evaluating $U_2$ in terms of its representation by the dual variable $N_5$, Eq.(\ref{U2}), for both models.

 The variation of $U_2$ versus $g_5$ from 0 to 1 occurs over the domain shrinking with $L \to \infty$ as the power $\sim L^{-1/\mu_5}$ where $\mu_5>0$ determines the scaling dimension $\Delta_5 = 2-1/\mu_5$ of the $\varphi^5$ term. 
[If $\Delta_5 <d=2$, this term is relevant and irrelevant otherwise].
Thus, $dU_2/dg_5 \propto L^{1/\mu_5} \to \infty$.
This derivative can be expressed in terms of averages of powers of $N_5$ with the help of the general relation for the derivative  $d \langle Q\rangle /dg_5 =g_5^{-1} [\langle Q N_5\rangle - \langle Q\rangle \langle N_5\rangle]$ of any quantity $Q$. This relation follows immediately from the representation (\ref{N5}) for both models. The result of the simulations for the truncated model are presented in Fig.~\ref{fig:dU2}.

The family of  the curves, Fig.~\ref{fig:dU2}, can be collapsed to a single master curve, Fig.~\ref{fig:masterg5},  by the scale factor $\lambda(L) \sim L^{-1/\mu_5}$ with
the exponent $\mu_5 = 0.534 \pm 0.008$. This exponent turns out to be consistent with the $\mu$-exponent of the 2D Ising model, $\mu=8/15\approx 0.533$, within 1-2\%
of the combined error -- systematic and statistical.  It is important to note that the range of $\lambda$ extends over almost 3 orders of magnitude.  In order to emphasize the quality of the collapse, we have included the plot Fig.~\ref{fig:deviation} showing two sizes $L=10,350$ rescaled to each other within a narrow range of $U^{-1}_2-1$. A visible deviation from scaling starts for $U^{-1}_2 -1 < 1$.  Similar behavior is demonstrated by the full model with $g_6=1$. Its master curve is also shown in Fig.~\ref{fig:dU2}, with the rescaling factor  characterized by the same exponent $\mu_5$. 

This concludes our analysis of the role of the symmetry breaking term $\varphi^5$ in 2D. Within the accuracy of 1-2\% and up to the simulated sizes of $L=350$ this term has the same scaling dimension as the linear one $\varphi$ in the Ising class. Using similar approach, higher odd terms can  be considered too. In response to the question \cite{SMA} about the role of the odd terms in the formal mapping \cite{Hubbard} of the LG critical point to the field theory, we conjecture that all odd terms have the same critical dimension of the field primary operator -- consistent with the Ising criticality.  This conjecture will be further supported in the section \ref{sec:dis}.

\section{LG criticality in 2D}\label{LG}
So far we have discussed the role of higher odd terms in the field theory along the line of the universality paradigm -- when a particular form of the action is not important as long as a system is close to the fixed point. The relation of this study to the actual LG criticality stems from the formal mapping of the classical gas to a field theory \cite{Hubbard}. 

Here we will analyze the LG transition in 2D gas of classical particles by simulating it directly. We choose the simplest  interacting potential -- the square well \cite{SW}. The NF method will be used to determine the critical behavior in this case too. 

The system of classical particles is described by the grand canonical  partition function
\beq
Y = \sum_{N=1}^\infty \frac{1}{N!}  e^{\tilde{\mu} N}\int d\vec{r}_1 ....d\vec{r}_N e^{- V},
\label{Y}
\eeq
where $V= (1/2)\sum_{ij} v(\vec{r}_i -\vec{r}_j)$ is the potential energy of binary interaction (normalized by temperature) between $N$ particles located at $\vec{r}_i, \, i=1,2,...N$ within the square area $L^2$ (now $L$ is a continuous length); $\tilde{\mu}$ is the chemical potential (normalized by temperature).

The interaction energy  $v(\vec{r})$ between two particles separated by a vector $\vec{r}$ is taken as the square well potential.
That is, $     v =   \infty$,  if $|\vec{r}|<\sigma$,  $ v= -\epsilon $, if $ \sigma \leq |\vec{r}|  \leq \tilde{\lambda}\sigma $, and $ v=0$, if $ r > \tilde{\lambda}\sigma$.
   Here $\sigma$ and $\tilde{\lambda}\sigma > \sigma$ are the hard and soft core radii, respectively, and $\epsilon >0$ characterizes attraction within the soft core shell. Since temperature is absorbed into the definition of $\epsilon$, we will be calling $1/\epsilon$ as "temperature" $T$ and $\tilde{\mu}$ as "chemical potential". Simulations have been conducted for $\lambda=1.5$. 

The quantities of interest are cumulants of the total number of particles $N$, that is, $\langle N^p \rangle$ with $p=1,2,3, ...$.
In the plane $(\tilde{\mu}, T)$ there is a line of Ist order phase transition between low and high density phases. This line ends by the critical point at some $\tilde{\mu}=\mu_c,\, T=T_c$. One of the significant difficulties in analyzing the LG transition is in finding this point in a controlled manner.   
Below, we will address this difficulty with the help of the NF method which leads to the critical point automatically -- along the same line as discussed in previous sections.  For this purpose we consider
the following Binder cummulant
\begin{equation}
	U_{4} = \frac{\langle (N-\langle N \rangle )^2 \rangle^2}{\langle (N-\langle N \rangle )^4 \rangle},
\label{U4}
\end{equation}
and its derivatives $dU_4/d\tilde{\mu}$ and $dU_4/d\epsilon$. [These derivatives can be expressed in terms of the cumulants $\langle N^p\rangle$, with $p=2,3, ...$  and $\langle N^p E\rangle$, where $E$ is the total energy of the system].

\begin{figure}
\vspace*{-0.5cm}
 \includegraphics[width=1.0 \columnwidth]{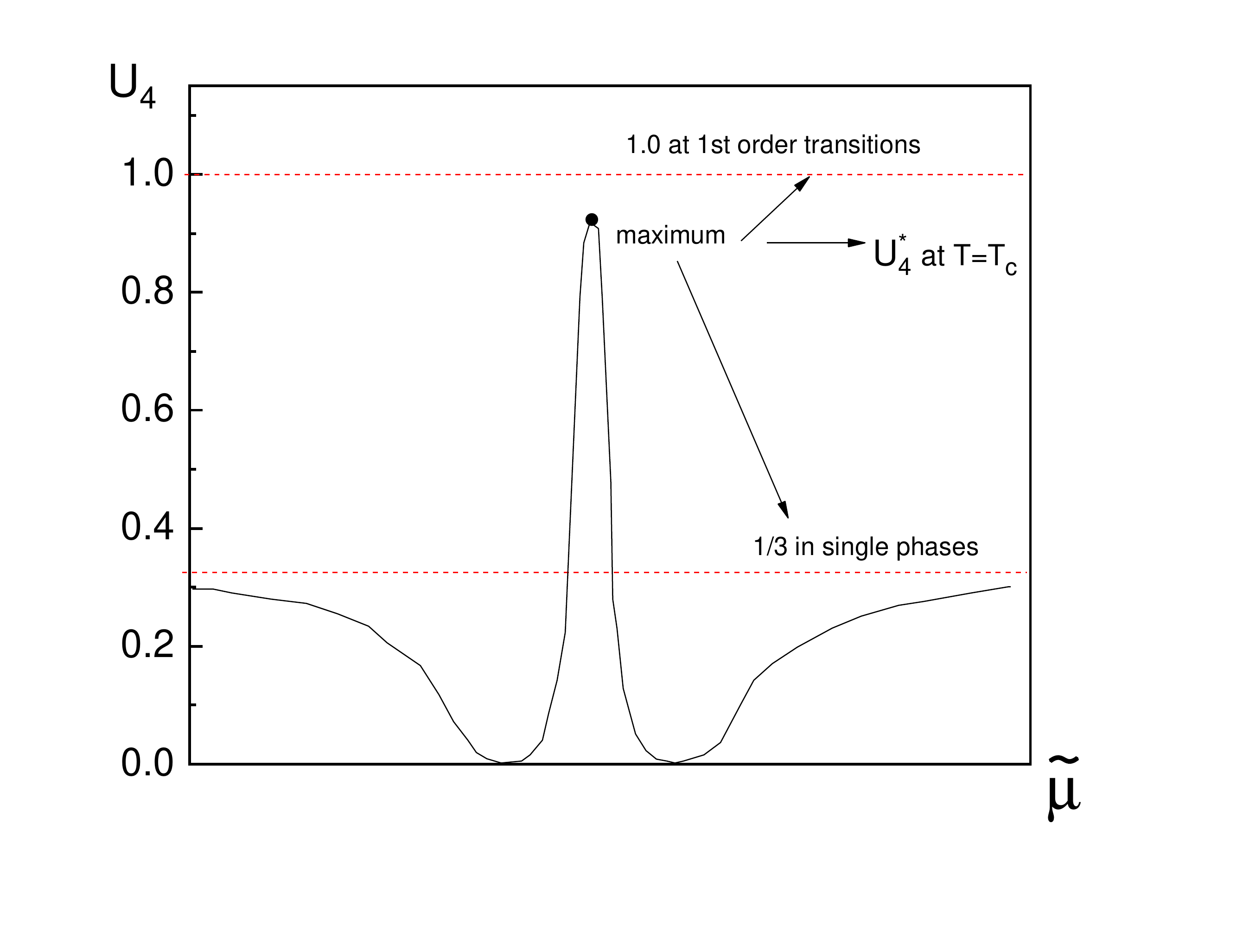}
\vskip-8mm
\caption{(Color online) Sketch of the Binder cumulant (\ref{U4})}
\label{fig:U4}
\vskip-5mm
\end{figure}

As discussed in Ref.~\cite{Binder}, this  cumulant has a specific form: away from the coexistence line it is $U_4 =1/3$ in the limit $L\to \infty$. At the coexistence line it has two dips corresponding to the densities of liquid and gas, with the peak in between corresponding to $U_4=1$. Above the critical point this maximum tends toward the value $U_4=1/3$. Thus, at the critical point the dips approach each other, with the peak reaching some intermediate value $1/3 <U_c <1$. This value is scale invariant \cite{Fisher_2003_2}. Fig.~\ref{fig:U4} illustrates this specific form of the cumulant.
\begin{figure}
\vspace*{-0.5cm}
 \includegraphics[width=1.0 \columnwidth]{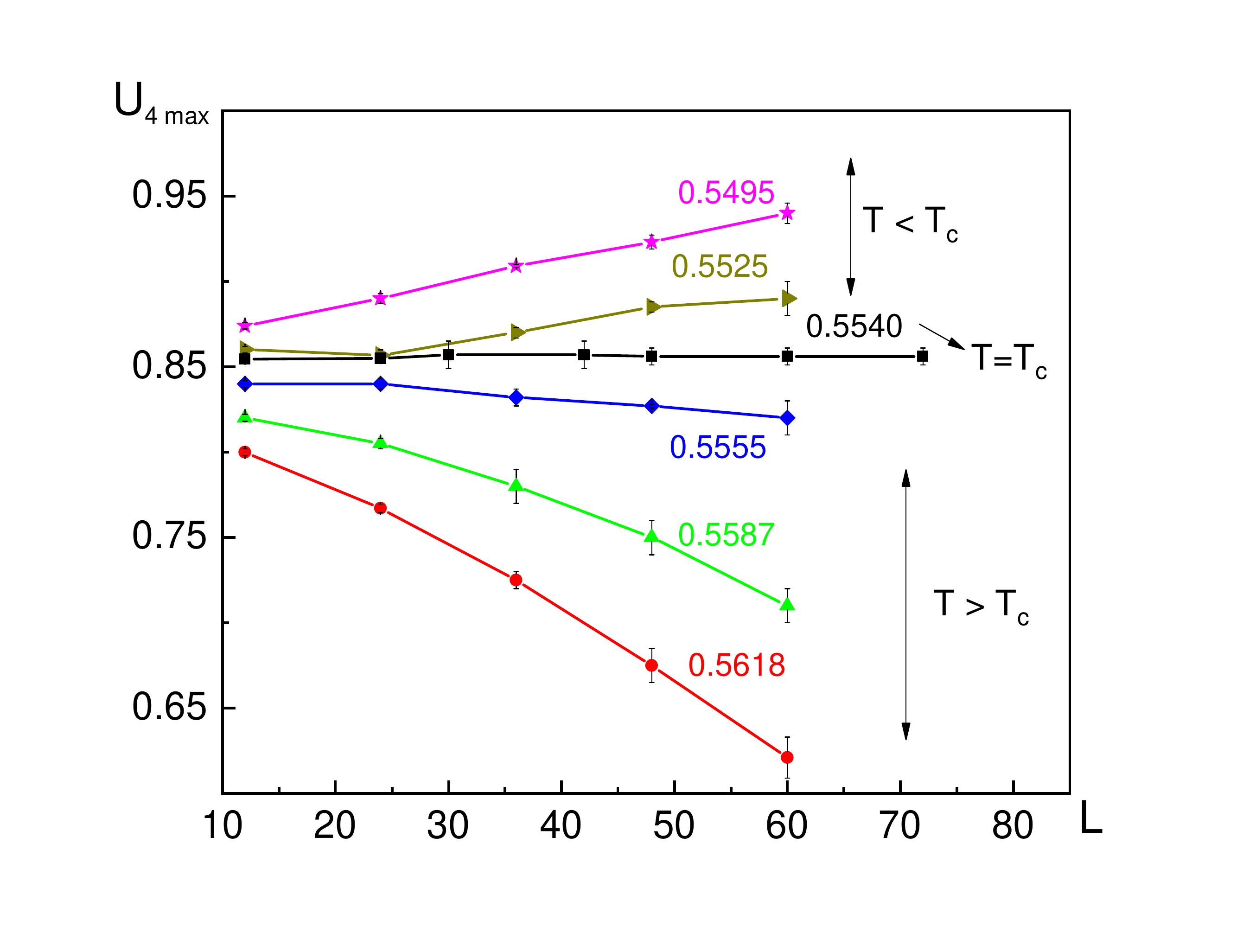}
\vskip-8mm
\caption{(Color online)
Monte Carlo results for the maximum of $U_4$ for several temperatures, $T$, (shown close to each curve) in regions $ T < T_c$ and $ T > T_c$. Lines are guides to eye.
The horizontal line, separatrix, determines the critical point, $T_c=0.5540 \pm 0.0005$
and $\tilde{\mu}_c=-3.700 \pm 0.005$.}
\label{fig:sep}
%\vskip-5mm
\end{figure}
\begin{figure}
\vspace*{-0.5cm}
 \includegraphics[width=1.0 \columnwidth]{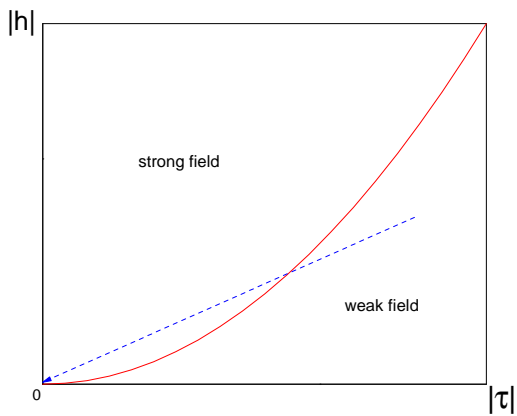}
\vskip-8mm
\caption{(Color online) A generic path (dashed line) toward the critical point ($h=0,\,\tau=0$) in presence of the mixing effect when $\mu <\nu$. The solid line, $h^* \sim |\tau|^{\nu/\mu}$  with  $\nu/\mu >1$, separates the regions of strong and weak field. }
\label{fig:generic}
%\vskip-5mm
\end{figure}
In other words, the critical point corresponds to the separatrix of the maximum of $U_4$ as a function of $T,\tilde{\mu}$ with respect  to $L \to \infty$. This suggests a protocol for finding the critical point: 1. choose some $T$ and find maximum of $U_4$ by adjusting $\tilde{\mu}$ for each size $ L$; 2. If this maximum flows toward 1 (toward 1/3), increase (decrease) $T$ and repeat the previous step  until the flow of $U_4$ maximum (versus $L$) saturates to some constant value. The result of this procedure is shown in Fig.~\ref{fig:sep}. It is important to emphasize that the accuracy of $T_c=0.5540\pm 0.0005$
and $\tilde{\mu}_c=-3.700 \pm 0.005$  is limited only by the maximum system size simulated and the numerical accuracy of $U_4$.
Obviously, no fitting procedure with respect to $T_c, \mu_c$ is required.

\begin{figure}
\vspace*{-0.5cm}
 \includegraphics[width=1.1 \columnwidth]{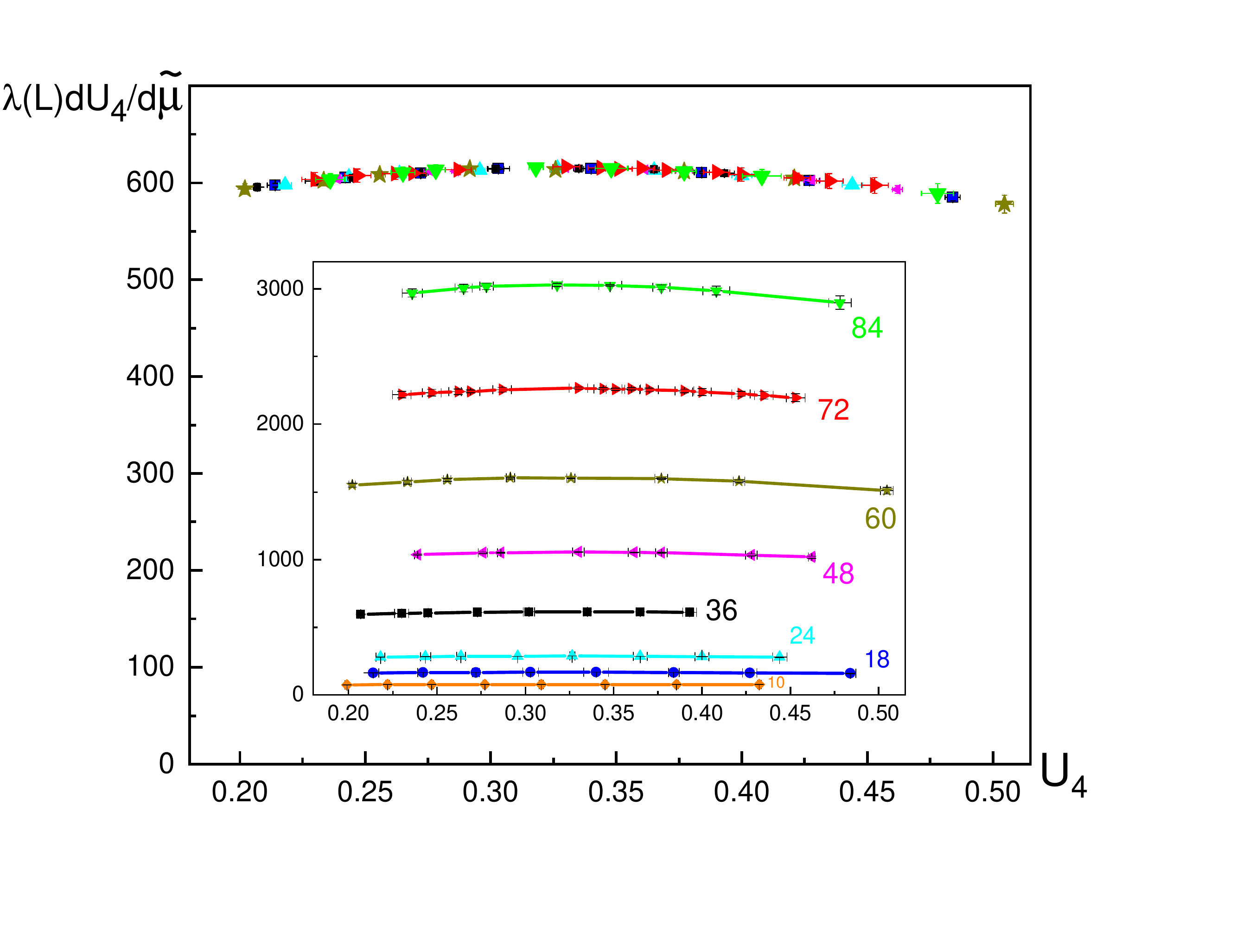}
\vskip-8mm
\caption{(Color online) 
The master curve obtained by "vertical" rescaling  of the data $dU_4/d\tilde{\mu}$ vs $U_4$ for sizes $L=10,...,84$ by the factor $\lambda(L)$ to achieve the best collapse. Inset: the original data for sizes shown close to each curve. Lines are guides to eye. }
\label{fig:dU4}
%\vskip-5mm
\end{figure}

\begin{figure}
%\vspace*{-0.5cm}
 \includegraphics[width=0.9 \columnwidth]{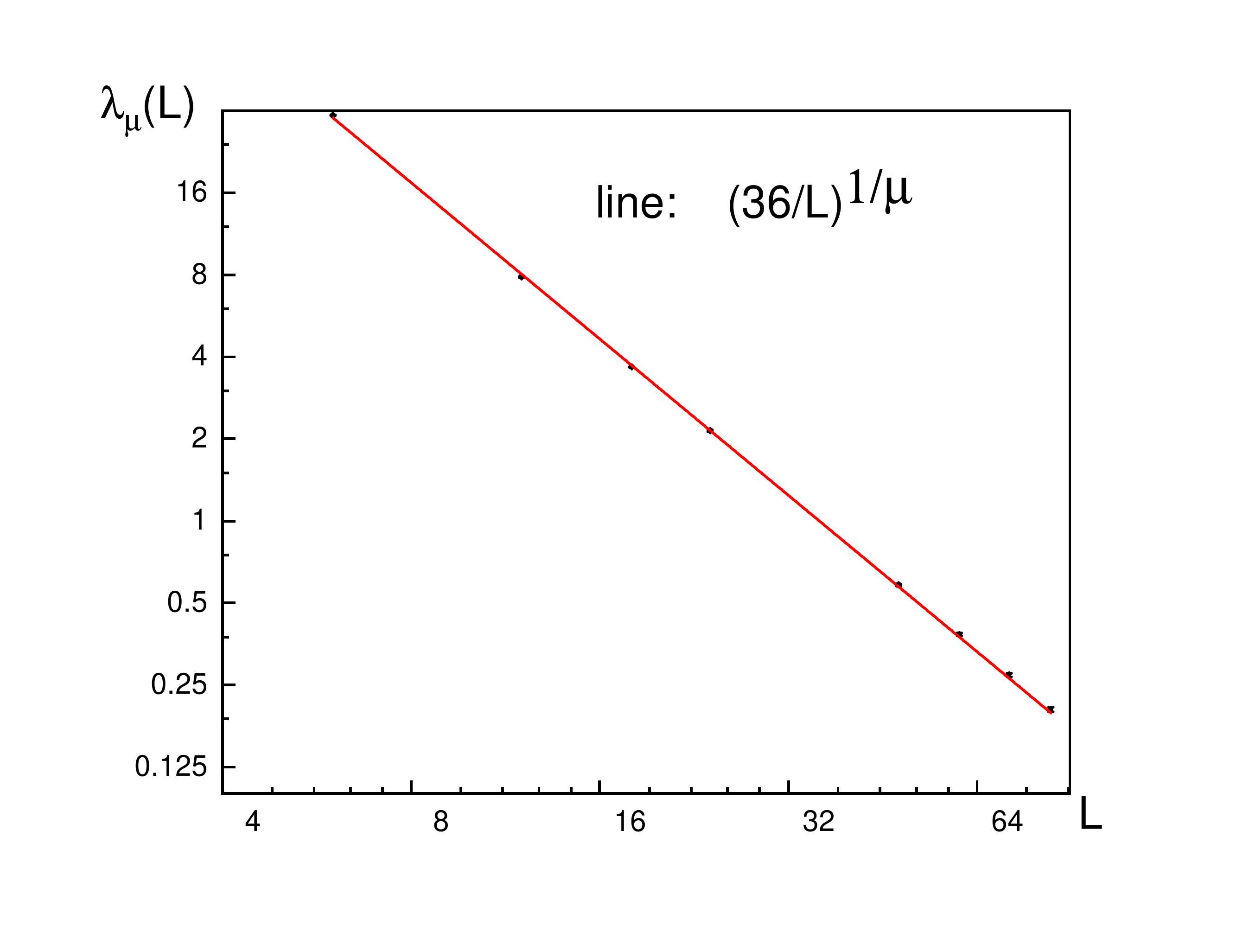}
%\vskip-8mm
\caption{(Color online) The rescaling factor $\lambda$ vs $L$ obtained from the data shown in Fig.~\ref{fig:dU4}. The value of the exponent $\mu=0.532 \pm 0.005$ is consistent with the Onsager value $\mu=8/15$. }
\label{fig:lamb_mu}
%\vskip-5mm
\end{figure}

Thus, while keeping $T=T_c$ the FSS can be conducted by tuning $\tilde{\mu}$ in the vicinity of $\mu_c$ so that
$U_4$ stays within the critical range $1/3 < U_4 <U_c$. Then, plotting $dU_4/d\tilde{\mu}$ versus $U_4$ should allow finding
the corresponding exponent.
There is one complication, though, -- a possibility of mixing of the primary operators in $N$ and $E$ in {\it a priory} unknown proportions as suggested in Refs.\cite{Mermin,Pokrovskii,Patashin} and further discussed in Refs.\cite{Fisher2000,Fisher_2001,Fisher_2003,Fisher_2003_2}. Thus, it is not known along which line in the space of the primary scaling operators  $(\tau,h)$ the system approaches criticality, if, say, $\tilde{\mu}$ is tuned toward $\mu_c$ while $T$ is kept  at its critical value $T=T_c$.
It is, however, possible to argue that, generically, the approach to the critical point should proceed along the line where the primary operator with smaller scaling dimension dominates. This argument goes as follows: the critical range can be divided into two parts -- of strong and weak field \cite{Landau}. The separation between the two regions are determined by the relation $h^*\sim \tau^{\nu/\mu}$ so that at $|h|>h^*$ the critical singularities are determined by $h$ rather than by $\tau \to 0$. Thus, if $\mu <\nu$, a generic path $\tilde{\mu} - \mu_c \sim r_1 \tau + r_2 h$ toward the critical point $\tau=0,h=0$  with non-zero mixing coefficients $r_{1,2}$ will belong to the region of strong field close enough to the critical point -- as sketched in Fig.~\ref{fig:generic}. Accordingly, conducting the FSS with respect to $\tilde{\mu}$ will give
the $\mu$ exponent. Conversely, if $\mu > \nu$, the approach should generically proceed along a path in the weak field region so that the
flowgram method will give the $\nu$ exponent.    The result of the fllowgram analysis of $dU_4/d\tilde{\mu}$ vs $U_4$
is shown in Fig.~\ref{fig:dU4} with the rescaling factor $\lambda$ plotted in Fig.~\ref{fig:lamb_mu}.
\begin{figure}
%\vspace*{-0.5cm}
 \includegraphics[width=1.0 \columnwidth]{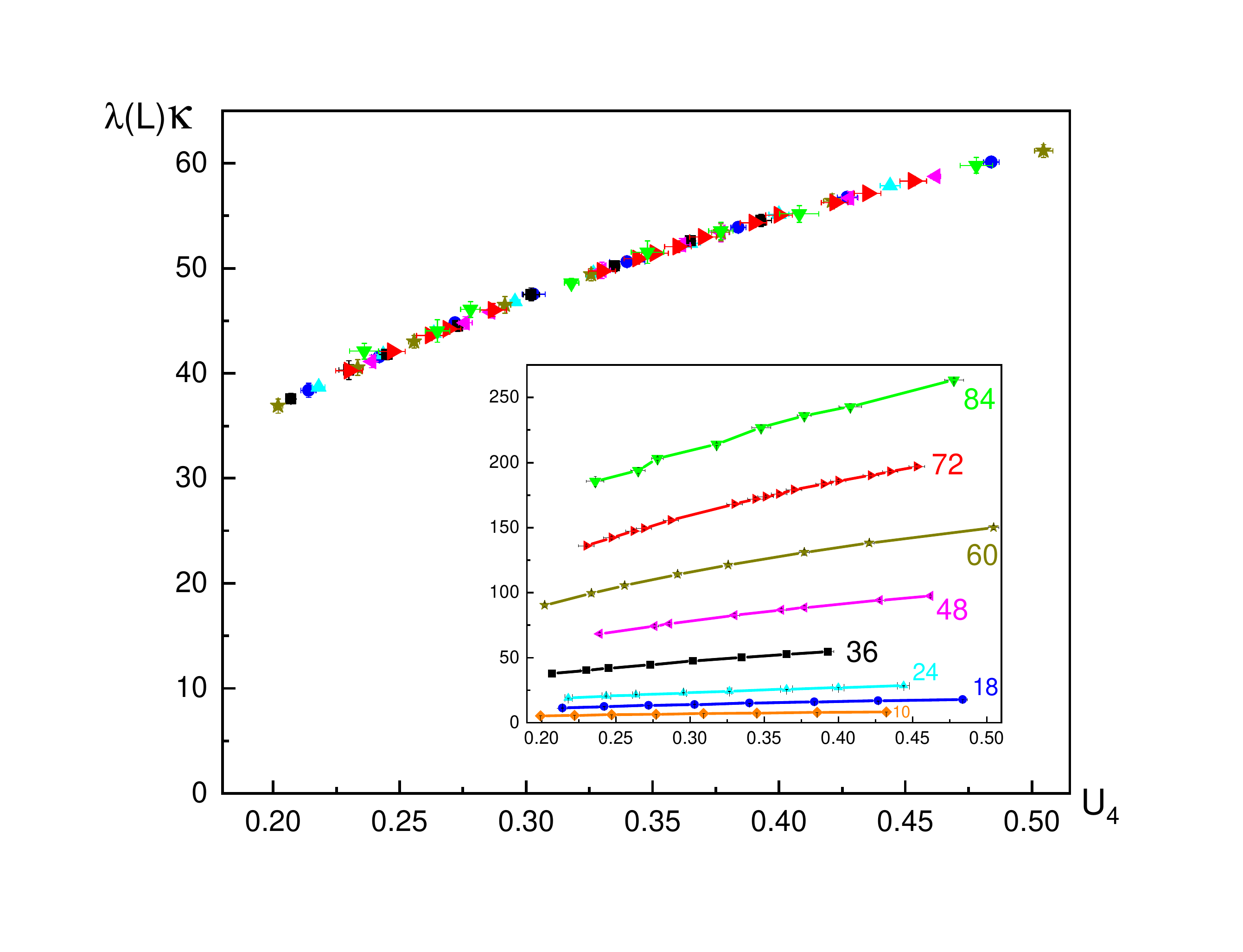}
%\vskip-8mm
\caption{(Color online) 
The master curve of the compressibility $\kappa$ vs $U_4$ obtained by rescaling of the data sets for various $L$. Inset: the original data for each size $L$ shown close to each curve. Lines are guides to eye.}
\label{fig:kappa}
%\vskip-5mm
\end{figure}
 \begin{figure}
%\vspace*{-0.5cm}
 \includegraphics[width=0.95 \columnwidth]{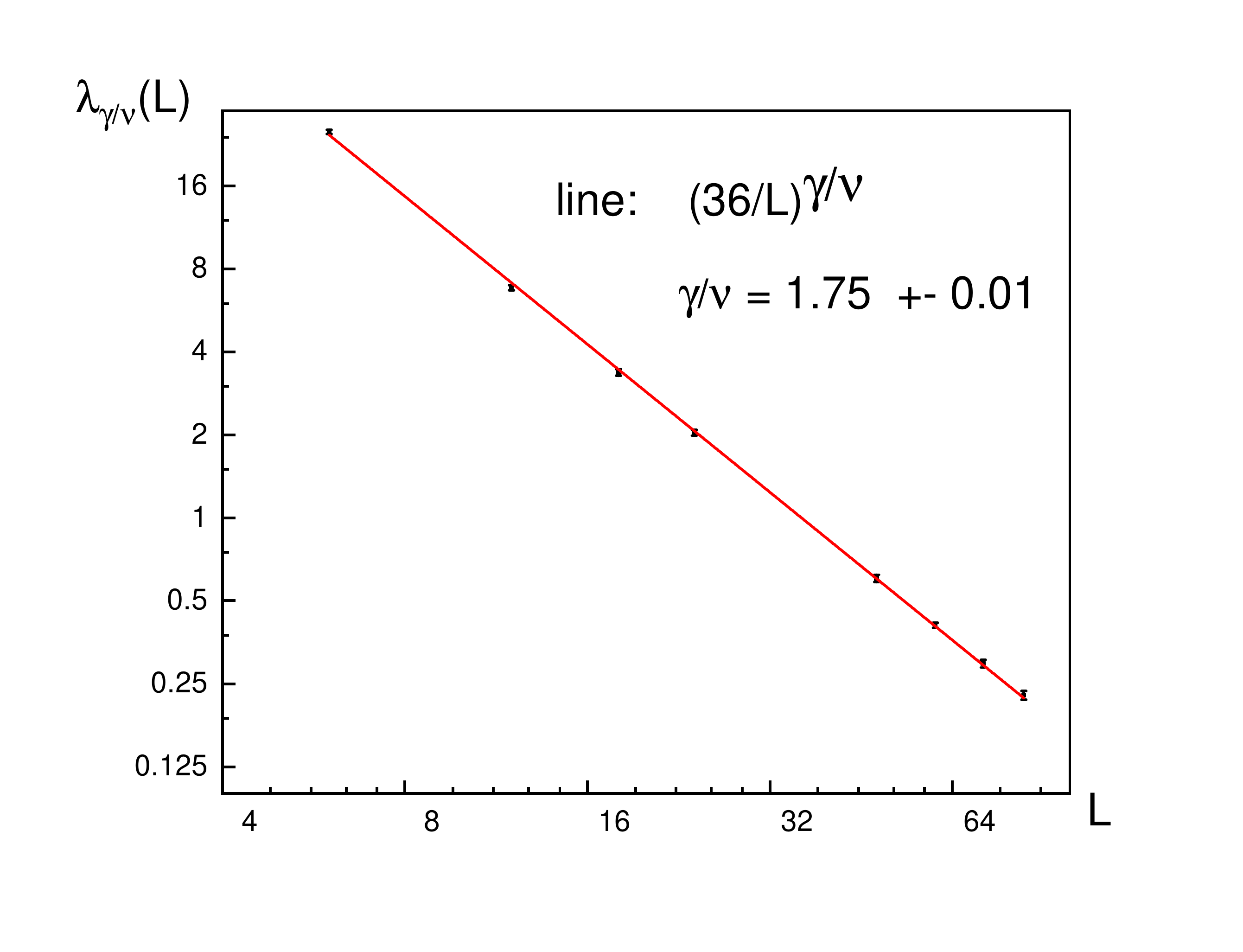}
%\vskip-8mm
\caption{(Color online) The rescaling factor $\lambda$ vs $L$ (symbols) obtained from the data shown in Fig.~\ref{fig:kappa}. The exponent $\gamma/\nu =1.75 \pm 0.01$ is consistent with the Onsager value $\gamma/\nu=7/4$. }
\label{fig:lam_k}
%\vskip-5mm
\end{figure}  
  As can be seen the resulting exponent $\mu=0.532 \pm 0.005 $ is consistent with the Ising value $8/15$ within 1\% of the statistical error. 

We have also analyzed the compressibility of the system $\kappa= \langle (N-\langle N\rangle)^2\rangle/L^2$ within the NF  method, that is, by plotting it vs $U_4$ in the critical range. The result is presented in Figs.~\ref{fig:kappa},\ref{fig:lam_k}.  
The found exponent $(1-1/\delta)/\mu=\gamma/\nu =1.75 \pm 0.05$, where $\delta,\, \gamma$ are the critical exponents (related to each other through the scaling relations),
is consistent with the Onsager value $7/4$. 
Thus, the results of our simulations strongly support the conjecture that the LG criticality in 2D belongs to the Ising class.

\section{Discussion and summary.}\label{sec:dis}
The NF method \cite{Annals,NJP} is a universal numerical tool in FSS analysis. Its main advantage comes from avoiding numerical fits where accurate knowledge of the position of the critical point is required -- in a strong contrast with the standard methods.  
The FSS relies on approaching the scaling regime where the role of the correlation length $\xi \sim |T-T_c|^{-\nu}$ is taken over by the system size $L$. Thus, the error of the universal scaling exponent $\delta \nu$ coming from the uncertainty $\delta T_c$ of the critical temperature $\delta \nu \approx \delta T_c/|T-T_c|\sim \delta T_c L^{1/\nu}$  grows with $L$. The situation becomes much worse in the case of the LG transition where the critical point is determined by two parameters -- critical temperature and pressure (or density). The NF method avoids this significant source of errors because no multi-parametric fits or extrapolations are used. As a result, its accuracy is solely determined by statistical errors of measuring appropriate Binder cumulants and their derivatives with respect to Hamiltonian parameters. 

 The scaling dimension of the $\phi^5$ term has been determined to be the same as of the linear term. We  
 conjecture that all odd terms are equivalent to the linear one at the criticality.   In simple terms, the mechanism can be illustrated by the following picture. A higher odd term $\sim \int d^dx \varphi^{2n+1}$ with $n=2,3,...$ in (\ref{H66}) can be decomposed as $\varphi=\varphi_L + \varphi_s$  into a long range part $\varphi_L$ and short range fluctuations $\varphi_s$ so that at the criticality the relevant contribution is $\sim \int d^dx \varphi_L \varphi_s^{2n}$ where     $\varphi_s^{2n}$ can be replaced by a short range contribution which is a non-critical constant.  

Here we give a qualitative argument in support of this conjecture using dual view on the field model. As can be seen from the dual representation (\ref{ZZ6}), any correlator $G_{2n+1}=\langle \varphi^{2n+1}(\vec{r}) \varphi^{2n+1}(0) \rangle $ with $n=0,1,2,...$ can be represented by a single loop of the bond integers $N_{ij}$ with only two open ends -- at $\vec{r}$ and at $0$. The logarithm of the statistical weight of such a loop depends on the loop structure and its length as an extensive value (with respect to $L$) at the criticality. The contribution to this weight depending on $n$  is finite and, thus, it cannot change the total weight in thermo-limit $L \to \infty$. This implies that all the correlators should be proportional to each other at the critical point. This argument shows that scaling dimensions of all odd terms determined with respect to the Ising fixed point should be identical (and equal to that of the linear term). Strictly speaking, however, this does not prove that these terms will not modify the criticality if these are added to the Hamiltonian.   
Here we proved only that $\varphi^5$ does not change the Ising universality. However, it is straightforward to apply the same
protocol for arbitrary odd term.

Here we have also addressed the LG criticality of a classical gas  in 2D free space. The analysis is based on applying the NF  method to the Binder cumulant showing a specific behavior \cite{Fisher_2003_2}. Our finding is that it is characterized by Onsager value of the critical exponent $\mu$. The same method can be used in 3D. However, the analysis is complicated by the low value of the exponent $\theta \approx 0.54$  determining correction to scaling. Thus, in order to suppress such corrections within the FSS analysis, much larger system sizes should be simulated. Alternatively, the fitting of the rescaling factor $\lambda (L)$ should involve two exponents -- the main one and $\theta$. This introduces significant uncertainty which requires  large computational efforts to minimize the contributions of errors from several fitting parameters.

A long standing problem in the theory of the LG criticality is the anomaly in the so called {\it diameter} -- the mean value of the liquid $n_l$ and gas $n_g$ densities along the liquid-vapor coexistence line. Absence of the underlying symmetry implies that the diameter must have a non-analytical term $\sim (T_c -T)^{1-\alpha}$ along the critical isochore (see in Ref.\cite{Landau}) with $\alpha$ being the heat capacity critical index. 
As suggested in Ref.\cite{Fisher2000} there should also be a much stronger term $ \sim (T_c -T)^{2\beta}$ where $\beta$ is the order parameter critical index. The attempts to observe this term directly \cite{Fisher_2001,Fisher_2003,Fisher_2003_2} were not very conclusive. The question, then, can be asked if the NF method can be used to resolve  the problem. Here we outline a path toward this goal.

We remind that the  heat capacity (in variables $\tilde{\mu},\, T$) diverges as $C \sim |T-T_c|^{-\alpha}$ along the coexistence line $\tilde{\mu}= \tilde{\mu}(T-T_c),\, T<T_c$  (cf. Ref.\cite{Landau}). On the sketch Fig.~\ref{fig:generic} this line is given by $h=0$. This divergence is much weaker than along a generic path (the dotted line in Fig.~\ref{fig:generic}) where $C \sim |T-T_c|^{-\gamma}$. In terms of the FSS, this means $C\sim L^{\alpha/\nu}$ and $C\sim L^{\gamma/\nu}$, respectively.  In 2D $\alpha/\nu =0$ and $\gamma/\nu=7/4$, and in 3D $\alpha/\nu \approx 0.2$ while $\gamma/\nu \approx 2$. This drastic difference in the divergence rate can be used to locate the coexistence line within FSS by measuring $C$ around the critical point (determined by the NF method as described above). Then, once $\tilde{\mu}$ is set along this line, the histogram of system density can be determined with the peaks $n_l$ and $n_g$ corresponding to the densities of liquid and gas. Since the critical density $n_c$ can be accurately determined by the NF method, the quantities $\eta_+ = n_l - n_c$ and $\eta_-= n_g - n_c$ can be identified with the order parameter values. Within FSS, these are characterized by $\eta_+ - \eta_- \sim L^{-\beta/\nu}$ and, if the anomaly $\tau^{2\beta}$, Ref.\cite{Fisher2000} is present, by $\eta_+ + \eta_- \sim L^{-2\beta/\nu}$. Within the NF method, these quantities should be plotted vs $U_4$ in its critical domain (collected also along the coexistence line) and then rescaled into two master curves with the corresponding values of the rescaling parameters $\lambda_+$, $\lambda_-$ for the sum and the difference, respectively. If the $2\beta$ anomaly is present, the log-log slope of $\lambda_+$ vs $L$ should be twice that of the slope of $\lambda_-$ vs $L$. It is worth mentioning that the outlined protocol does not involve the direct fitting of $\eta_+ + \eta_-$ by $\sim (T_c -T)^{2\beta}$. This project will be discussed elsewhere.

%At this point we also mention that evaluating the derivative of $U_4$ along the coexistence line (the line $h=0$ in the sketch Fig.~\ref{fig:generic}) and plotting it versus $U_4$ as described in Sec.\ref{LG} would allow to measure the $\nu$ exponent. This derivative should be characterized by the rescaling parameter $\lambda \sim L^{1/\nu}$ (instead of $\lambda \sim L^{1/\mu}$).      

Summarizing, the numerical flowgram method has been applied to the problem of LG criticality in 2D and the critical correlation length exponent $\mu$ has been determined  to be consistent with 2D Ising class within the combined error of 1-2\%. The main advantage of the method is that it does not require the accurate knowledge of  the position of neither the critical point nor the coexistence line. Instead, these quantities follow as a byproduct of the method. The role of the odd terms in the real scalar field theory near the critical point has been addressed too in the context of the general mapping of the LG transition to the field theory. The analysis of the $\varphi^5$ term revealed that its critical dimension is the same as that of the linear term $\varphi$. We have put forward a  conjecture that in 2D all odd terms have the same critical dimension. This excludes the possibility of non-Ising LG criticality.

\noindent {\it Acknowledgments}.
We acknowledge helpful discussions with Victor Bondarev, Alexander Patashinski, David Schmeltzer and Aleksey Tsvelik.  This work was supported by the National Science Foundation under the grants PHY1314469 and DMR1720251.

\end{document}